\documentclass[letterpaper,twocolumn,10pt]{article}
\usepackage{usenix2019_v3}

\usepackage{tikz}
\usepackage{amsmath}
\usepackage{booktabs}
\usepackage{graphicx}

\newcommand{\stitle}[1]{\vspace*{0.4em}\noindent{\bf #1.\/}}

\newcommand{\squishlist}{
	\begin{list}{$\bullet$}
		{ \setlength{\itemsep}{1pt}
			\setlength{\parsep}{1pt}
			\setlength{\topsep}{2.5pt}
			\setlength{\partopsep}{0.5pt}
			\setlength{\leftmargin}{1em}
			\setlength{\labelwidth}{1em}
			\setlength{\labelsep}{0.6em}
		}
}
\newcommand{\squishend}{\end{list}}

\begin{document}

\date{}

\title{SARIA: Persistence-Aware KV Caching with Coordinated DRAM--SSD Reload for LLM Agents}

\author{
\begin{minipage}{0.95\textwidth}
\centering
{\normalfont
Hui Li$^{1}$, Qihui Zhou$^{2}$, Xiao Yan$^{3}$, Shumeng Zhang$^{4}$, Zijie Dai$^{1}$\\[0.5em]
Zequn Gong$^{2}$, Mingcong Song$^{2}$, Guoping Long$^{2}$, Hongjie Si$^{2}$, James Cheng$^{1}$\\[0.8em]
}
{\itshape
$^{1}$The Chinese University of Hong Kong \quad $^{2}$Huawei \quad $^{3}$Wuhan University \quad $^{4}$Xiamen University
}
\end{minipage}
}

\maketitle

\begin{abstract}
Multi-turn LLM agents usually reuse long context prefixes, making key value (KV) caching crucial for reducing prefill latency. SSDs provide a low-cost extension to limited DRAM capacity but introduce two key challenges.
Eagerly persisting all KV cache blocks incurs substantial write traffic and rapidly consumes finite SSD endurance.
Meanwhile, the small number of SSDs provisioned in typical GPU servers limits aggregate reload bandwidth, making SSD-resident KV cache much slower to retrieve than that in DRAM and inflating prefill latency.

We present \textsc{Saria} (Selective Admission and Reload for Inference Acceleration), a KV caching system that jointly manages DRAM and SSD by exploiting the session-level temporal locality of LLM agents to reduce SSD writes and accelerate cache reload.
On the write path, \textsc{Saria} uses eviction and reuse history to selectively persist KV cache blocks that exhibit reuse opportunities beyond their DRAM residency, avoiding unnecessary writes for KV cache already hosted in DRAM.
On the read path, \textsc{Saria} promotes selected SSD-hit KV cache blocks from GPU memory into DRAM after prefill, allowing subsequent reuse within the same session to reduce SSD read traffic while limiting DRAM contention.
A transfer coordinator paces DRAM reads while SSD transfers are active, allowing both paths to make progress and efficiently share GPU ingress bandwidth.

We implement \textsc{Saria} on LMCache MP and evaluate it using workloads from TraceLab and WEKA, covering both full-attention and hybrid-attention models.
The results show that \textsc{Saria} reduces SSD writes by 57.1--58.9\% compared with popular full duplicate caching and the mean time to first token (TTFT) by 42.3--79.7\% over DRAM-only caching.
\end{abstract}

\section{Introduction}
\label{sec:introduction}

\begin{figure}[!t]
  \centering
  \includegraphics[width=0.99\linewidth]{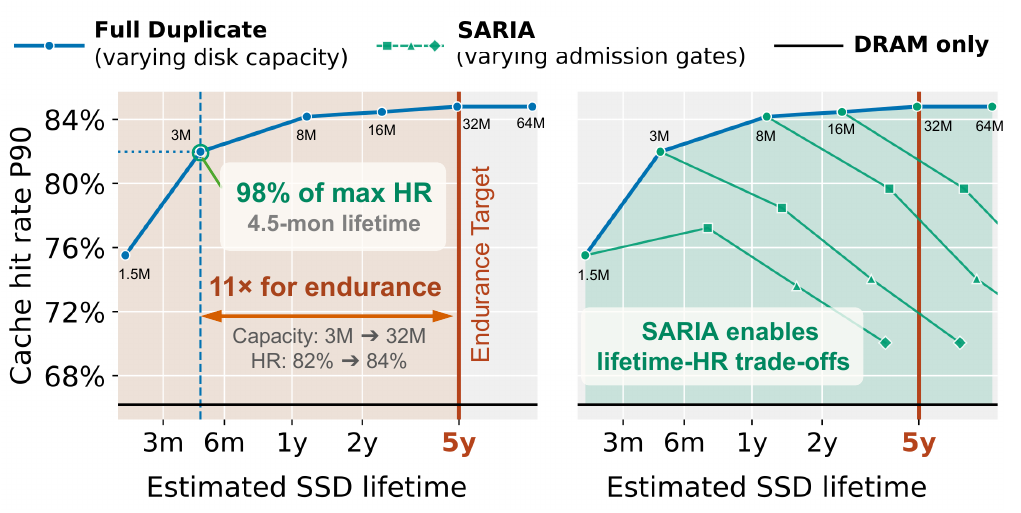}
  \caption{
  P90 cache hit rate and estimated SSD lifetime for GLM-5.2 (1~DWPD) as SSD capacity varies (labels denote token slots, e.g., 1.5M). SARIA (right) can extend SSD lifetime relative to full duplicate caching (left) at the same capacity by trading cache hit rate for longer lifetime.
  }

  \label{fig:alpha-fig}
\end{figure}

Agentic LLM services are turning the key value (KV) cache from ephemeral, per-request state into reusable serving state that spans multiple interaction turns. System prompts, tool schemas, repository instructions, and conversation history often constitute long prefixes that recur throughout an agent session. Retaining the KV cache of these prefixes allows subsequent requests to avoid repeated prefill computation, reducing time to first token (TTFT)~\cite{cachedattention,pensieve,marconi,cachewise,tracelab}. However, GPU HBM and host DRAM provide limited capacity at high cost, making it difficult to retain all reusable KV cache blocks.

SSDs offer a storage tier with larger capacity and lower cost. Existing systems such as Mooncake~\cite{mooncake} and LMCache~\cite{lmcache} have already extended KV cache hierarchies from
host DRAM to SSDs.
Meanwhile, I/O technologies such as GPUDirect Storage (GDS)~\cite{nvidiaGDS} enable direct SSD-to-GPU data transfer at high efficiency.
With sufficient SSD parallelism and an optimized I/O stack, recent work shows that SSD-based KV caching can approach DRAM performance~\cite{tutti}.
These advances highlight the potential of SSD-based KV caching but two challenges need to be tackled when deploying SSD caching on typical GPU servers.

\stitle{\textit{Challenge 1: Finite SSD endurance}}
SSDs can only sustain a finite cumulative write volume before exhausting their rated endurance budget~\cite{micronSSDEndurance}. Existing systems duplicate all newly generated KV blocks from DRAM to SSD, which we call \emph{full duplicate caching}. As such, write traffic scales with the model's KV footprint and prefill throughput. For example, vLLM reports 12,963 prefill tokens/s per GPU for GLM-5.2 with 8K-token inputs on B300 GPUs~\cite{vllmGLM52B300}. At this rate, each GPU would generate approximately 1,145~MiB/s of KV writes, or 103.7~TB/day. A 4-TB SSD rated at one drive write per day (DWPD) would consume its five-year write budget in about 70 days of continuous operation.

Increasing SSD capacity provides a larger write budget, but the capacity needed for endurance can far exceed that needed for caching. For Figure~\ref{fig:alpha-fig}, we replay 350 TraceLab sessions following the setup described in \S\ref{sec:evaluation-setup}. Increasing the cache size from 3M token slots to 32M token slots (the capacity required for five-year endurance) only boosts cache hit rate from 82.0\% to 84.8\%. As such, using larger SSDs  is not economical in terms of the performance gain.


\stitle{\textit{Challenge 2: Limited SSD parallelism}}
The number of SSDs on GPU servers is usually too small to match GPU ingress bandwidth. For example, AWS's \texttt{p5.48xlarge} provides 8 H100 GPUs and 8 local NVMe devices~\cite{awsP5}, whereas at least 4 x4 PCIe SSDs are needed to match one x16 GPU link at the same PCIe generation~\cite{nvidiaGDS}. Adding more drives consumes bays and PCIe resources and increases cost. With limited device parallelism, SSD bandwidth remains a bottleneck when reloading the cached prefixes.

Existing SSD-based KV cache systems adopt two reload paths. The first fetches SSD-resident KV cache directly into GPU memory via GDS~\cite{flexkvGDSReload, dynamoKVBMReload}.
Since this path bypasses DRAM, DRAM copies are not created for SSD hits, a behavior we refer to as \emph{zero promotion}. Repeated SSD hits therefore remain bound by SSD bandwidth.
The second path loads SSD-resident KV cache into DRAM before transferring it to the GPU~\cite{sglangHiCache}.
These systems retain all reloaded blocks in DRAM for subsequent reuse, which we refer to as \emph{full promotion}. Although these copies reduce subsequent SSD reads, the promoted KV cache blocks compete with other active sessions for DRAM.


\stitle{Our System \textsc{Saria}} To tackle the two challenges, we present \textsc{Saria} (Selective Admission and Reload for Inference Acceleration), a DRAM--SSD KV caching system for agent workloads.
The central idea is to guide SSD persistence and promotion using session reuse history. \textsc{Saria}'s \emph{persistence-aware SSD admission} preserves KV cache blocks whose reuse extends beyond DRAM residency, while \emph{lazy partial DRAM promotion} and \emph{coordinated multi-source reload} combine GDS reads with DRAM replicas to accelerate subsequent reloads. Together, they reduce SSD writes and reload latency under limited SSD capacity and parallelism.

To account for finite SSD endurance, \textsc{Saria} uses repeated DRAM evictions as evidence that a block needs SSD persistence.
This is because requests within the same session often reuse KV cache blocks while they remain in DRAM, but contention from other active sessions or long inter-request intervals can evict these blocks before they are reused. Therefore, repeatedly evicted KV cache blocks have high potential for reuse after DRAM eviction.
We also persist exceptionally popular KV cache blocks (based on their reuse counts) before they are evicted.
\textsc{Saria} allows users to trade off SSD write volume against cache coverage by configuring eviction and reuse-count thresholds (\S\ref{sec:design-admission}).
\textsc{Saria} further designs unified context metadata to support both classical full-attention models and emerging hybrid-attention models and plans valid prefix reloads from both DRAM and SSD (\S\ref{sec:design-prefix}).

To tackle limited SSD parallelism, \textsc{Saria} combines direct GDS reloads with lazy partial promotion to accelerate subsequent cache reuse.
After a request reloads SSD-hit KV cache into GPU memory through GDS and completes prefill, lazy promotion copies some SSD-hit blocks from HBM into DRAM. We choose a promotion ratio so that DRAM and SSD can fully utilize the GPU's ingress bandwidth while avoiding DRAM capacity competition.
Subsequent requests can then reload different portions of the KV cache from DRAM and SSD concurrently to reduce SSD traffic (\S\ref{sec:design-promotion}).

Loading KV cache concurrently from both DRAM and SSD requires careful coordination for good performance. We observe that DRAM transfers can consume most of the GPU ingress bandwidth and leave a long SSD-only completion tail. As such, \textsc{Saria}'s coordinator paces DRAM transfers while GDS is active and only restores full DRAM bandwidth after SSD transfer finishes. To fully utilize GPU ingress bandwidth, \textsc{Saria} also uses the residual GPU ingress bandwidth to serve DRAM reloads for other requests (\S\ref{sec:design-coordinator}).

We implement \textsc{Saria} on LMCache MP and evaluate it using TraceLab and WEKA workloads, covering full-attention and hybrid models and single- and dual-SSD configurations. The results show that the selective admission of \textsc{Saria} reduces SSD write volume by 57.1--58.9\% compared with the full duplicate baseline. The mean TTFT is also reduced by 42.3--79.7\% over DRAM-only caching. The ablation study shows that partial promotion, tier overlap, and coordination together reduce the cache reload time by 26.7--31.9\% for requests in the slowest quartile.


To summarize, we make the following contributions:

\squishlist

\item We identify short SSD endurance from eager persistence  and limited reload bandwidth from insufficient SSD parallelism as two key challenges for SSD-based KV caching.

\item We design \textsc{Saria} with \emph{Persistence-Aware SSD Admission}, which uses eviction and reuse history to reduce SSD writes while preserving cache beyond DRAM residency. To mitigate limited SSD parallelism, \textsc{Saria} combines direct GDS reloads with partial DRAM promotion and coordinated multi-source transfers to reduce reload latency.

\item We evaluate \textsc{Saria} across full-attention and hybrid models under varying cache pressure and SSD configurations, quantifying its write--performance trade-off and the benefits of its reload mechanisms through end-to-end experiments and ablations.

\squishend

\begin{figure}[t]
  \centering
  \includegraphics[width=0.99\linewidth]{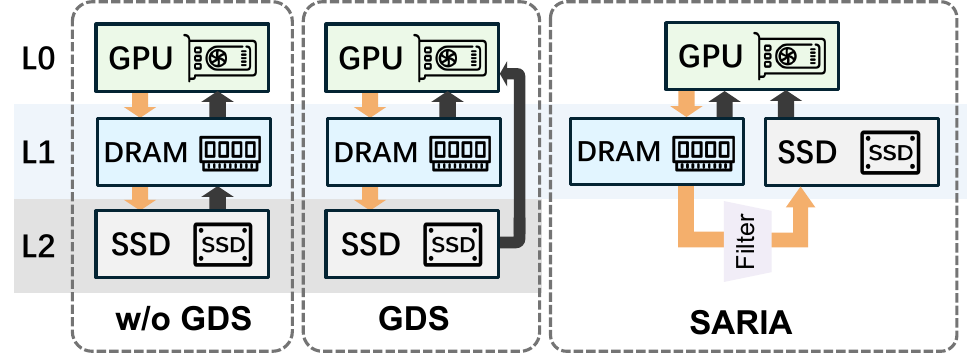}
  \caption{
  KV cache data path comparison among DRAM staging (w/o GDS), GDS, and \textsc{Saria}. \textsc{Saria} selectively persists KV cache to SSD and enables direct reloads from both DRAM and SSD. Orange and black arrows denote offload and reload paths, respectively.
  }
  \label{fig:GDS-compare}
\end{figure}

\section{Background and Challenges}
\label{sec:background}

\subsection{Hierarchical KV Caching}
\label{sec:background-hierarchy}
\label{sec:background-hybrid}

KV caching allows later requests to restore cached prefixes and prefill only unmatched suffixes.
CachedAttention~\cite{cachedattention}, Mooncake~\cite{mooncake}, and LMCache~\cite{lmcache} extend the KV cache from GPU memory to host DRAM and SSD.
DRAM provides fast reloads, while SSD extends capacity. SSD reloads either stage through DRAM or use direct I/O such as SPIN~\cite{spin} and GPUDirect Storage (GDS)~\cite{nvidiaGDS}.
The GDS paths in FlexKV~\cite{flexkvGDSReload} and Dynamo KVBM~\cite{dynamoKVBMReload} reload SSD blocks without retaining DRAM copies (\emph{zero promotion}); HiCache's DRAM-staged path supports \emph{full promotion}, retaining all reloaded blocks for subsequent requests~\cite{sglangHiCache}. Tutti shows that SSD caching can approach DRAM-backed performance with sufficient SSD parallelism and optimized I/O~\cite{tutti}.

Different model architectures impose different requirements on KV cache access and reuse.  Full-attention (FA) KV can be reused at any matched prefix boundary. Hybrid models combining FA with recurrent layers, such as linear attention (LA), also require a checkpoint at that boundary because later states cannot generally be rolled back~\cite{marconi,unifiedRadixCache}.  Their access patterns therefore differ: FA blocks can serve prefix hits ending at different positions, while a recurrent checkpoint is accessed when computation resumes at its specific boundary.

\subsection{Endurance-Constrained SSD Provisioning}
\label{sec:background-ssd}

Repeated program/erase cycles wear NAND flash, limiting SSD write endurance. Manufacturers rate endurance in terabytes written (TBW) or drive writes per day (DWPD), the daily write volume normalized by capacity~\cite{micronSSDEndurance}. We consider standard-endurance TLC SSDs rated at 1 DWPD over five years~\cite{kioxiaCD8PR}. A $C$-TB SSD therefore has a $1{,}825C$-TB host-write budget, which a sustained $B$-GB/s write stream consumes in $1{,}825C/(86.4B)$ days.

Full duplicate caching persists every newly computed KV block, so write traffic scales with the computed token rate and KV bytes per token. Table~\ref{tab:model-cache-profile} uses published per-H200 KV generation rates to estimate a 4-TB SSD's lifetime and the capacity needed to sustain five years of writes. Qwen3-235B and GLM-5.2 use FP8 KV; Qwen3-30B-A3B uses BF16 KV with 4K-token inputs and batch size 32~\cite{qwen3Model,glm52Model,qwen330BConfig}. For MiniMax-M2.5, we use the reported KV throughput at 32K-token inputs directly.

\begin{table}[t]
  \centering
  \small
  \setlength{\tabcolsep}{2pt}
  \caption{Per-H200 SSD endurance requirements under continuous full duplicate caching, assuming a five-year, 1-DWPD rating. The 4-TB lifetime is the time to exhaust the rated host-write budget; minimum capacity sustains the same write rate for five years.}
  \label{tab:model-cache-profile}
  \label{tab:ssd-endurance}
  \begin{tabular*}{\columnwidth}{@{\extracolsep{\fill}}lrrr@{}}
    \toprule
    Model & \shortstack{KV/s\\(MiB/s)} & \shortstack{4-TB SSD\\lifetime (d)} & \shortstack{Min. capacity\\for 5 yr (TB)} \\
    \midrule
    Qwen3-30B-A3B~\cite{sglangHPCOpsPrefill} & 6,515 & 12.4 & 590.2 \\
    MiniMax-M2.5~\cite{prefillAsAService} & 893 & 90 & 80.9 \\
    Qwen3-235B~\cite{moePrefill} & 383 & 210 & 34.7 \\
    GLM-5.2~\cite{sglangGLM52Prefill} & 181 & 445 & 16.4 \\
    \bottomrule
  \end{tabular*}
\end{table}

At 1 DWPD, sustaining $B$~GB/s per GPU for five years requires at least $C_{\min}=86.4B$~TB of aggregate SSD capacity per GPU. Full duplicate caching can therefore require substantially more SSD capacity for write endurance than for effective caching.


\subsection{Limited SSD Reload Bandwidth}
\label{sec:background-underprovisioned}

Approaching the GPU's PCIe ingress bandwidth with SSD-backed KV reload requires multiple drives operating in parallel.  A typical NVMe SSD uses an x4 PCIe link, while a PCIe-attached GPU typically uses x16.  At the same PCIe generation, at least four typical x4 NVMe SSDs must operate in parallel to match one x16 GPU link~\cite{nvidiaGDS}.

Dense GPU servers commonly provide fewer drives than needed to match aggregate GPU ingress bandwidth. A common configuration in mainstream GPU servers pairs eight GPUs with eight local NVMe SSDs~\cite{hammerspaceTier0}. For example, AWS's \texttt{p5.48xlarge} has eight H100 GPUs and eight local NVMe devices~\cite{awsP5}; GIGABYTE's eight-H200 configuration supports at most eight hot-swap PCIe 5.0 NVMe bays~\cite{gigabyteG593SD1}, while Lenovo's eight-B200 configuration supports at most sixteen NVMe bays~\cite{lenovoSR680aB200}. Adding drives consumes bays and PCIe resources and increases cost, while shared links can limit aggregate bandwidth~\cite{nvidiaGDS}. Long-prefix reloads can therefore remain bandwidth-bound.

\section{Observations and Motivations}
\label{sec:observations-and-motivation}
To characterize session-level KV reuse under cache pressure, we analyze the same 350 reconstructed TraceLab sessions used in our evaluation~\cite{tracelab,tracelabDataset}, averaging 18.4 requests per session and 83.7K prompt tokens per request. Section~\ref{sec:evaluation-setup} describes workload construction; Table~\ref{tab:workloads} summarizes the 24-hour replay.

To avoid distorting multi-turn arrival patterns by scaling inter-turn times~\cite{servegen}, we preserve each session's request sequence and inter-request intervals, varying only session start times. We place starts within 12-, 4-, and 2-hour windows, denoted lower, medium, and higher load, respectively. Shorter windows make more active session contexts compete for the same DRAM capacity, raising cache pressure without changing within-session reuse timing.

\subsection{KV Reuse under Cache Pressure}
\label{sec:observation-cache-pressure}

\begin{figure}[t]
  \centering
  \includegraphics[width=\linewidth]{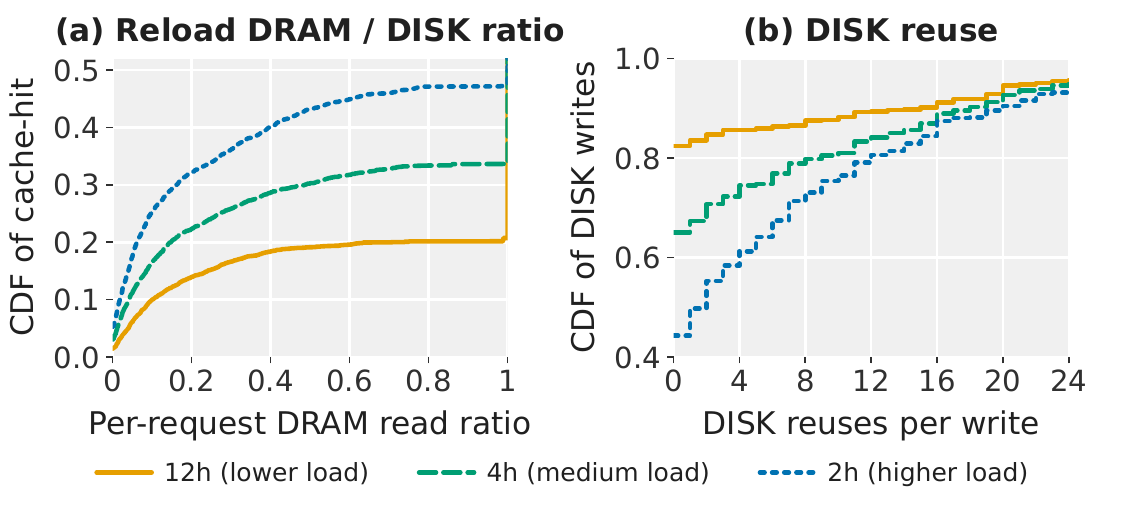}
  \caption{GLM-5.2 reuse under full duplicate caching: (a) DRAM read ratios among cache-hit requests.  (b) subsequent SSD reads per write.
  }
  \label{fig:cache-pressure-reuse}
\end{figure}


We observe that requests within an agent session often reuse overlapping context over short intervals. Such temporal locality allows DRAM to serve much of the reuse as long as the corresponding KV cache remains resident.
We therefore examine how cache pressure affects the role of SSD persistence, focusing on how much reusable KV cache can already be served from DRAM and how often an SSD copy is actually read after being written.

To quantify these effects, we simulate full duplicate caching under the three load settings. We model a 100~GB DRAM cache for GLM-5.2 using approximately 2.2 million token slots, based on its FP8 KV footprint. SSD capacity is sufficient to avoid eviction, and SSD reads do not repopulate DRAM.

Figure~\ref{fig:cache-pressure-reuse}(a) shows the fraction of cache-hit traffic served by DRAM. In the lower-load 12-hour setting, 79.3\% of cache-hit requests obtain all their cached blocks from DRAM, which supplies 76.9\% of total cache-read bytes. As load increases, more active sessions compete for the fixed DRAM capacity, causing reusable blocks to be evicted earlier. In the higher-load 2-hour setting, these fractions fall to 52.4\% and 44.2\%, respectively, substantially increasing reliance on SSD.

Figure~\ref{fig:cache-pressure-reuse}(b) measures the value of individual SSD writes by counting how many subsequent SSD reads each persisted block receives through the end of the replay. Under lower load, 82.5\% of SSD writes receive no subsequent SSD read. This fraction decreases to 65.0\% in the 4-hour setting and 44.3\% in the higher-load 2-hour setting. Thus, under lower cache pressure, DRAM already serves much of the observed reuse, leaving many eagerly persisted SSD copies unused. As cache pressure increases, however, reusable blocks are more likely to outlive their DRAM residency, making SSD persistence increasingly valuable.

\noindent\textbf{Observation 1.} \emph{Session-level temporal locality allows DRAM to serve most KV reuse under lower cache pressure, leaving many SSD writes unused; as cache pressure increases, more reuse extends beyond DRAM residency and becomes dependent on SSD persistence.}

\subsection{DRAM Competition under Full Promotion}
\label{sec:observation-promotion-competition}

\begin{figure}[t]
  \centering
  \includegraphics[width=\linewidth]{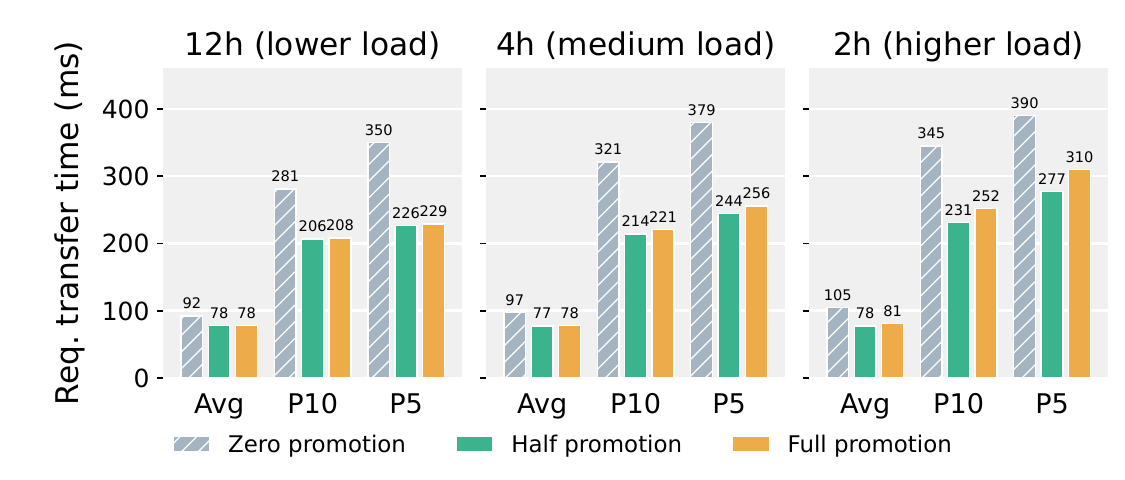}
  \caption{GLM-5.2 request transfer time under zero, half, and full promotion (lower is better). P10 and P5 average the slowest 10\% and 5\% of requests, respectively. Half promotion improves tail transfer time over full promotion, with a larger advantage at higher load.}
  \label{fig:promotion-transfer-time}
\end{figure}

The session-level temporal locality observed above also creates an opportunity on the reload path. After an SSD-resident block is reloaded through GDS, retaining a copy in DRAM can accelerate subsequent reuse by reducing reliance on limited SSD bandwidth. However, promoted replicas consume additional DRAM capacity and compete with the KV cache of active sessions. We therefore examine whether promoting more SSD-hit KV cache always reduces transfer time as cache pressure increases.

We compare zero, half, and full promotion under the three load settings, holding the workload and all other cache settings fixed within each setting. Transfer times are measured under ideal overlap between DRAM and SSD transfers, with GDS bandwidth set to half the GPU-ingress bandwidth. Half promotion uses a fixed target of 50\% DRAM coverage of the currently reused prefix: it creates only enough replicas to reach that target, counting blocks already in DRAM. Figure~\ref{fig:promotion-transfer-time} reports mean request transfer time and the mean times for the slowest 10\% and 5\% of requests, denoted P10 and P5, respectively.

Both half and full promotion reduce mean transfer time relative to zero promotion, but full promotion does not consistently improve on half promotion. Under lower load, both achieve a mean of approximately 78~ms, with similar P5 times of 226~ms for half promotion and 229~ms for full promotion. The tail gap widens as load increases: under medium load, half promotion reduces P5 from 256 to 244~ms; under higher load, it reduces P10 from 252 to 231~ms and P5 from 310 to 277~ms, while also lowering the mean from 81 to 78~ms. Thus, full promotion retains much of the average benefit but delivers smaller tail improvements at higher load than half promotion.

The additional replicas also increase DRAM replacement. In the 2-hour replay, cumulative DRAM block evictions rise from 59,589 with zero promotion to 97,097 with half promotion and 175,314 with full promotion. More aggressive promotion therefore creates more replicas but also evicts more resident KV cache blocks, potentially removing reusable blocks before their next access. Together with the transfer-time results, these counts indicate that limiting promotion can alleviate DRAM competition and improve tail transfer performance under cache pressure.

\noindent\textbf{Observation 2.} \emph{Session-level temporal locality makes DRAM promotion beneficial, but aggressive promotion intensifies DRAM competition under cache pressure. Reducing the promotion ratio can alleviate this competition and improve tail transfer latency.}

\section{SARIA}
\label{sec:design}

\subsection{System Overview}
\label{sec:design-overview}

\textsc{Saria} manages a hierarchical DRAM--SSD KV cache for serving agent workloads under limited SSD endurance and reload bandwidth.  It selectively persists KV cache using reuse and eviction history accumulated in DRAM, and maintains context metadata to apply admission consistently across hybrid KV types and recover prefixes distributed across storage tiers.  When a request reloads SSD-backed KV cache, lazy partial promotion creates selected DRAM replicas for subsequent reuse, with the promotion volume jointly determined by the reuse recency of DRAM eviction candidates and the DRAM-to-SSD bandwidth ratio.  A coordinator schedules the resulting DRAM and GDS transfers to efficiently share GPU ingress bandwidth.

\begin{figure}[t]
  \centering
  \includegraphics[width=0.99\linewidth]{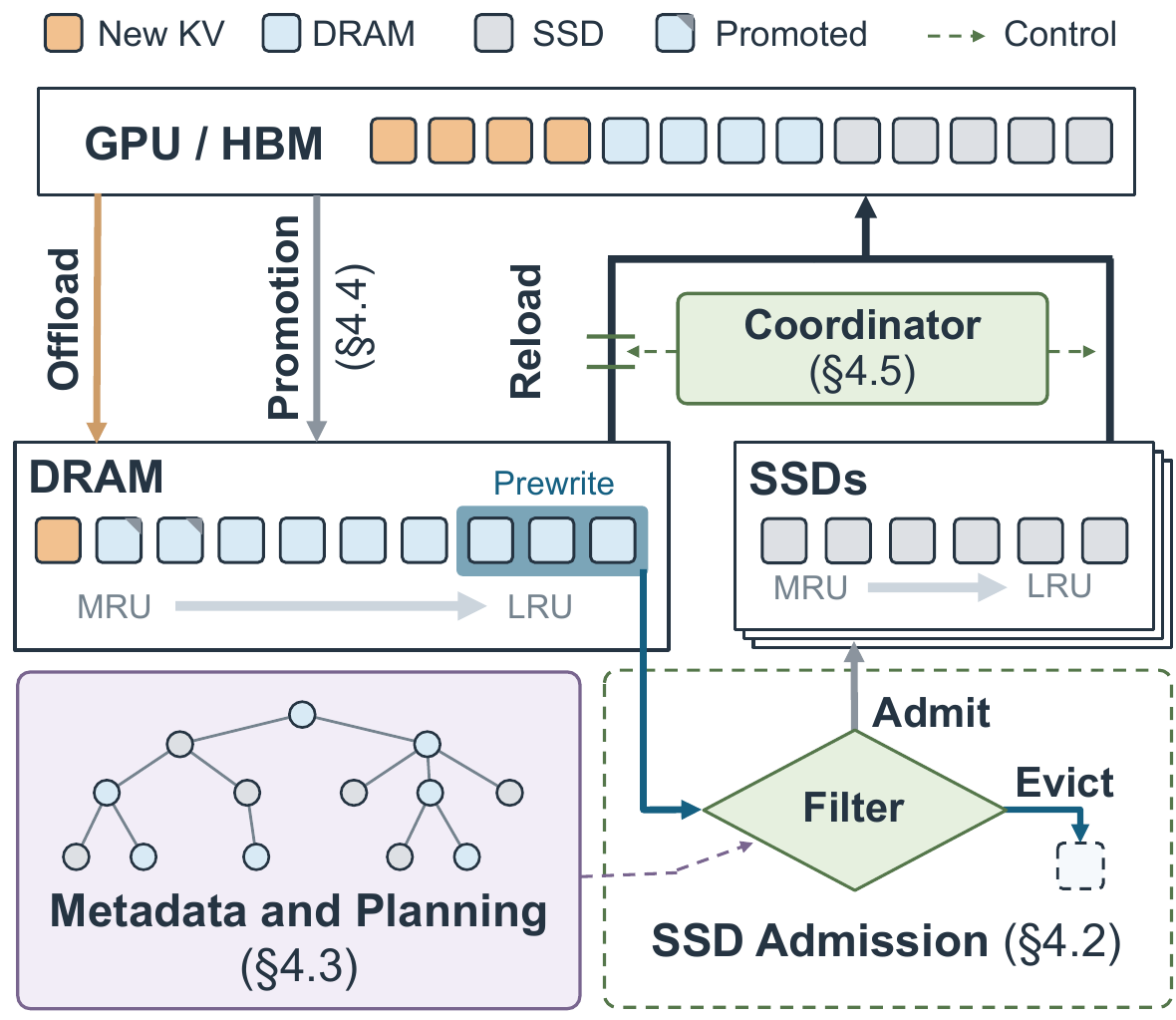}
  \caption{Overview of \textsc{Saria} architecture and data paths.
  }
  \label{fig:pact-kv-architecture}
\end{figure}

Figure~\ref{fig:pact-kv-architecture} illustrates how \textsc{Saria} manages KV cache throughout a request's lifecycle. Newly generated KV cache blocks are first placed in DRAM.
\emph{Persistence-Aware SSD Admission} then uses accumulated reuse and eviction history to determine which blocks should be persisted to SSD for reuse beyond their DRAM residency (\S\ref{sec:design-admission}).
Unified context metadata preserves this history across DRAM residencies and tracks the current location of each KV block.
During lookup, the cache manager uses this metadata to construct an ordered reload plan for the reusable prefix, which may span both DRAM and SSD (\S\ref{sec:design-prefix}).
After prefill completes, \emph{Lazy Partial DRAM Promotion} copies selected SSD-hit blocks from HBM into DRAM to accelerate subsequent reuse (\S\ref{sec:design-promotion}).
Finally, the transfer coordinator uses each request's reload plan to pace DRAM reads alongside GDS transfers, allowing the two paths to efficiently share GPU ingress bandwidth (\S\ref{sec:design-coordinator}).

\subsection{Persistence-Aware SSD Admission}
\label{sec:design-admission}

Every SSD write consumes a finite endurance budget, even if the persisted KV cache is never read from SSD (\S\ref{sec:background-ssd}).  Observation~1 (\S\ref{sec:observation-cache-pressure}) shows that DRAM supplies most cached reuse under normal load and many eagerly persisted copies provide no read benefit: in the reference replay, 82.5\% of SSD writes under full duplicate caching receive no subsequent SSD read.
This motivates an admission filter that defers SSD persistence while DRAM serves reuse and seeks evidence that retaining the KV cache beyond its DRAM residency would be beneficial.

A natural admission signal is request popularity.
HiCache, for example, uses access counts to filter HBM-to-DRAM backups~\cite{sglangHiCache}, but does not make a separate admission decision for SSD endurance.
For predominantly session-private agent context, however, these counts largely describe reuse within a session and do not reflect the DRAM competition from other sessions.
Increased competition from other active sessions can shorten a block's DRAM residency even when its own access pattern remains unchanged.
Moreover, a cumulative access count does not distinguish reuse concentrated within a short burst from reuse spread over longer intervals.
Blocks with similar popularity can therefore have very different needs for SSD persistence.

\begin{figure}[t]
  \centering
  \includegraphics[width=\linewidth]{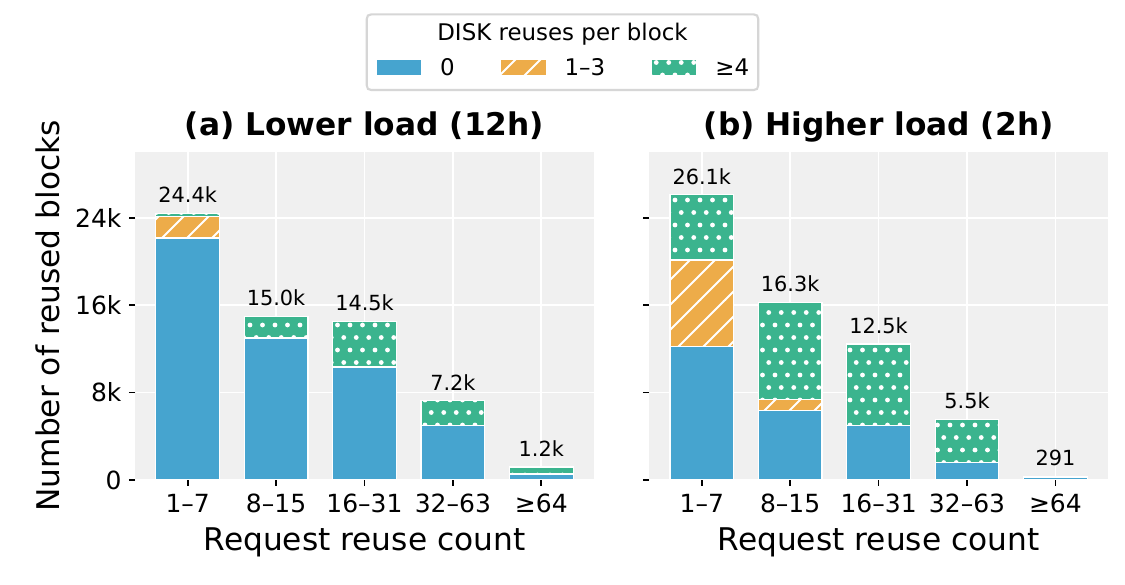}
  \caption{GLM-5.2 SSD reuse by request popularity under 12-hour and 2-hour session-start windows.
  }
  \label{fig:admission-popularity}
\end{figure}

Figure~\ref{fig:admission-popularity} illustrates this mismatch by grouping reused blocks from the 12-hour and 2-hour settings of Observation~1 according to their request-reuse counts.
Each bar separates blocks with zero, one to three, or at least four SSD reads.
Among blocks reused 16--31 times, the fraction receiving at least one SSD read increases from 29.0\% under lower load to 60.2\% under higher load.
Even among blocks reused at least 64 times, 44.5\% receive no SSD read in the 12-hour setting.
Request popularity alone therefore does not capture the need for SSD persistence under cache competition.


\textsc{Saria} uses repeated DRAM evictions as a practical signal of persistence demand.  A KV block's first eviction indicates that DRAM can no longer retain it under the current workload. If the block is subsequently reused and later reaches another eviction boundary, its history includes reuse that DRAM could not cover.
This cross-residency history incorporates the competition experienced by that block without explicitly estimating its reuse interval or DRAM retention time.
Repeated eviction does not guarantee another return, but provides direct evidence that SSD persistence could have protected useful KV cache across an earlier eviction. We additionally retain a request-reuse threshold so that exceptionally popular blocks can qualify for persistence before experiencing repeated evictions.

For each logical context block $b$, \textsc{Saria} tracks its DRAM-eviction count $E_b$ and request-reuse count $R_b$ across DRAM residencies.  The eviction counter advances when the block is selected at a DRAM eviction boundary.  A block becomes eligible for SSD persistence when
\begin{equation}
  \operatorname{admit}(b) =
  \left[E_b \geq \theta_E\right]
  \lor
  \left[R_b \geq \theta_R\right].
  \label{eq:persistence-admission}
\end{equation}
The two thresholds expose a trade-off between SSD write traffic and persistence coverage.
Larger thresholds filter more writes but delay protection, potentially requiring recomputation on early returns; smaller thresholds admit KV cache blocks sooner at greater write cost.  Once persisted, a block can serve subsequent reuse after DRAM eviction for as long as it remains on SSD.

Admission determines \emph{whether} to persist a block; write scheduling determines \emph{when} persistence begins.
When an eligible, non-persisted block reaches the DRAM eviction path, \textsc{Saria} writes it to SSD before reclaiming its DRAM storage. The DRAM copy remains available as the write source until the transfer completes.
Persistence can also begin earlier.
The reuse path can initiate an asynchronous write when $R_b$ reaches $\theta_R$.  Prewrite evaluates candidates using $E_b+1$, allowing blocks that would qualify at their next eviction to start writing while still resident in DRAM.
When sufficient lead time and write bandwidth are available, these asynchronous writes move SSD traffic off the DRAM reclamation critical path.

\subsection{Unified Metadata and Reload Planning}
\label{sec:design-prefix}

Applying admission independently to physical blocks can leave a hybrid model without the KV cache needed to recover a context prefix.  Full-attention (FA) KV through a position $c$ remains useful when later requests extend beyond $c$, whereas a linear-attention (LA) recurrent state at $c$ is reused as a checkpoint at that particular boundary~\cite{unifiedRadixCache,marconi}.  These components can therefore accumulate different access and eviction histories despite representing the same context position.  Independent admission may therefore persist FA blocks without the LA checkpoint needed to reuse that prefix.  Selective persistence and independent replacement also distribute reusable blocks across DRAM and SSD, producing interleaved sources such as DRAM--SSD--DRAM. Recovery must therefore establish a valid prefix boundary and assign a source to every cache object required at that boundary.

\begin{figure}[t]
  \centering
  \includegraphics[width=\linewidth]{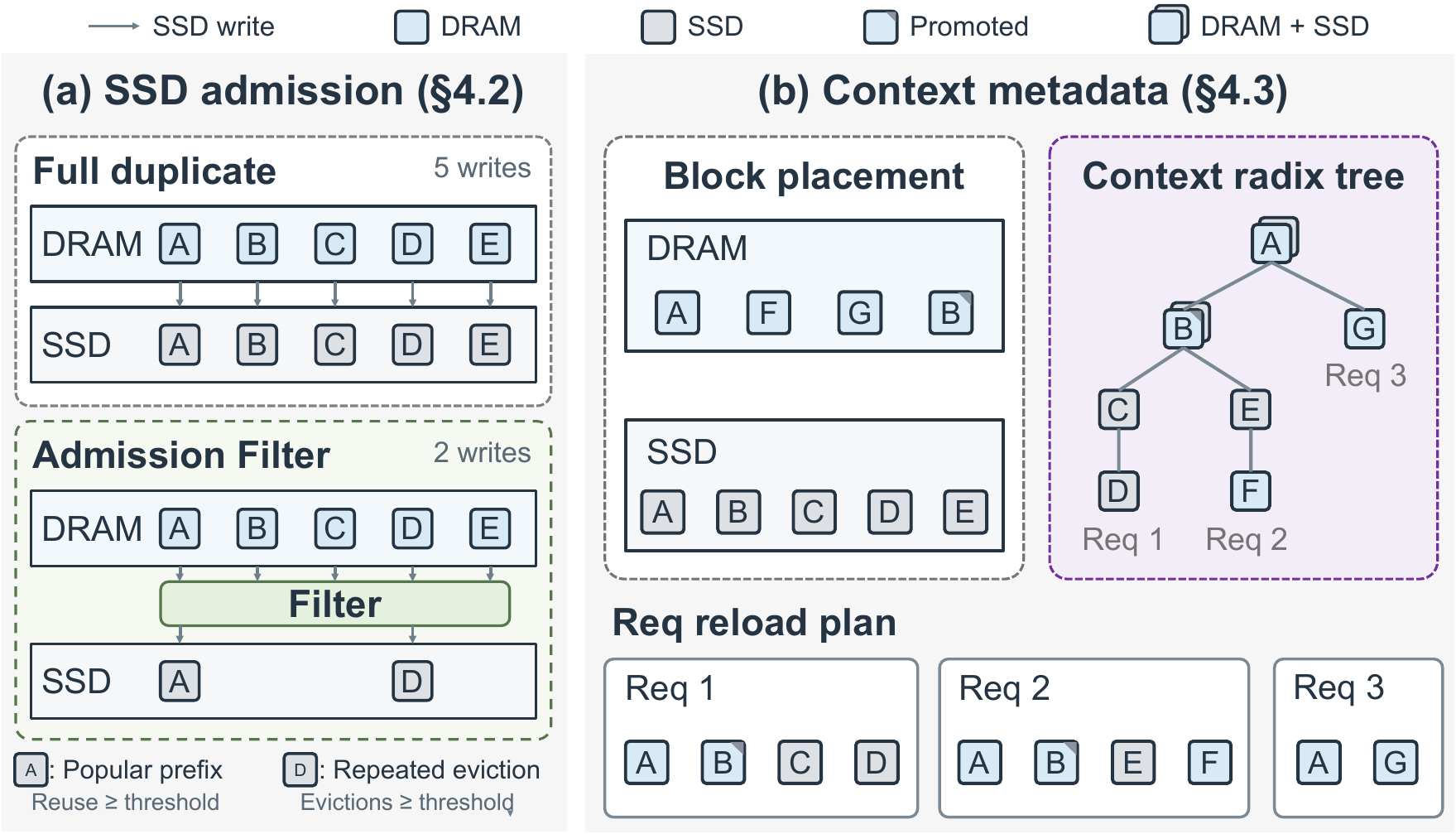}
  \caption{
  (a) Persistence-aware SSD admission. (b) Unified metadata and mixed-source reload planning.
  }

  \label{fig:admission-metadata}
\end{figure}

\textsc{Saria} organizes context metadata in a radix tree, following the prefix organization used in HiCache and Unified Radix Cache~\cite{sglangHiCache,unifiedRadixCache}.  Each node represents a block-aligned context prefix and associates its physical cache objects with shared reuse and eviction counts and per-object storage locations.  Admission is evaluated at this context granularity and applied to the associated objects.  This allows a checkpoint to share the admission evidence accumulated by its context instead of requiring every component to independently reach the threshold.  Metadata is retained across evictions so that returning blocks preserve their history across residencies.  We manage metadata with a sliding time window, reclaiming a block's metadata only after an entire window passes without reuse or recomputation.

At lookup, the cache manager combines the KV cache available in DRAM and SSD into a unified reload plan (Figure~\ref{fig:admission-metadata}(b)).
The plan identifies the reusable prefix and selects each required object from an available DRAM or SSD source, preserving block order even when hits alternate between tiers.

\subsection{Lazy Partial DRAM Promotion}
\label{sec:design-promotion}

\begin{figure}[t]
  \centering
  \includegraphics[width=0.8\linewidth]{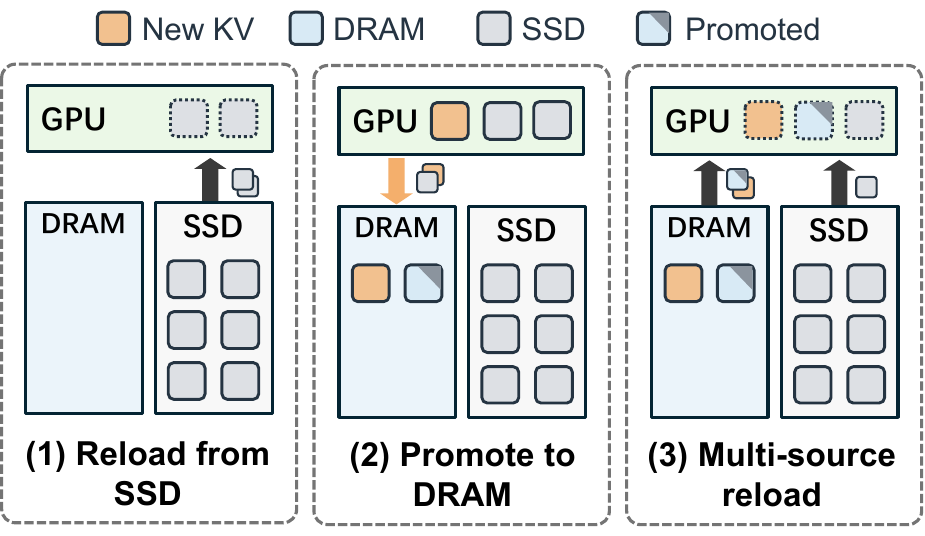}
  \caption{
  Illustration of the \emph{Lazy Partial DRAM Promotion} workflow.
  }
  \label{fig:promotion-fig}
\end{figure}

Limited SSD parallelism constrains GDS reload bandwidth (\S\ref{sec:background-underprovisioned}).  Observation~2 (\S\ref{sec:observation-promotion-competition}) shows that promotion reduces mean transfer time but increases competition for DRAM capacity. Limiting the promotion ratio can ease this competition and reduce tail transfer latency.
\textsc{Saria} therefore creates DRAM replicas on demand while limiting their volume. After a request reloads KV cache from SSD through GDS and completes prefill, \textsc{Saria} copies selected SSD-hit blocks from HBM into DRAM (Figure~\ref{fig:promotion-fig}).
If these replicas remain resident, subsequent requests can reuse them from DRAM instead of repeatedly relying on the bandwidth-limited SSD path.

\textsc{Saria} limits the DRAM competition introduced by promotion along two dimensions.
First, it avoids displacing recently reused KV cache blocks in DRAM merely to accommodate promoted replicas.
A candidate block $b$ may be evicted for promotion only if $t_{\mathrm{now}}-t_b^{\mathrm{reuse}}\geq P_{\tau}$, where $t_b^{\mathrm{reuse}}$ is its last reuse time and $P_{\tau}$ is a recency threshold.
Recently reused candidates indicate strong competition for DRAM capacity and are protected from promotion-induced eviction.
\textsc{Saria} therefore promotes only as many blocks as can be accommodated by candidates satisfying this threshold.
When few candidates qualify, the promotion volume is reduced, potentially to zero.

Second, \textsc{Saria} bounds the target promotion volume according to the relative bandwidth of the DRAM and SSD reload paths.
Since DRAM and GDS can serve different portions of a prefix concurrently, DRAM only needs to supply the bandwidth not provided by SSD to saturate the shared GPU ingress bandwidth.

Let $V$ denote the total bytes of a subsequent reload, $B_{\mathrm{disk}}$ the sustained GDS-to-GPU bandwidth, and $B_{\mathrm{GPU}}$ the shared GPU-ingress bandwidth, where $B_{\mathrm{disk}}<B_{\mathrm{GPU}}$.
Assuming DRAM can sustain the residual bandwidth $B_{\mathrm{GPU}}-B_{\mathrm{disk}}$ while GDS is active, the target fraction of reload bytes served from DRAM is
\begin{equation}
  r^\star = 1 - \frac{B_{\mathrm{disk}}}{B_{\mathrm{GPU}}}
  \label{eq:ideal-promotion-ratio}
\end{equation}
Serving $r^\star \cdot V$ bytes from DRAM and $(1-r^\star)\cdot V$ bytes from SSD  allows the two paths to finish together while fully utilizing GPU ingress, yielding a reload time of $V/B_{\mathrm{GPU}}$.
Promoting beyond this fraction consumes additional DRAM capacity without further reducing reload time.

This bandwidth balance guides the target DRAM share $r\in[0,1]$, which also gives the block fraction for equal-sized blocks.
If a request currently reloads $D_0$ blocks from DRAM and $S_0$ blocks from SSD, the number of additional SSD-backed blocks targeted for promotion is
\begin{equation}
  P_{\mathrm{target}} =
  \max\left(0, r(D_0 + S_0) - D_0\right)
  \label{eq:promotion-target}
\end{equation}
Blocks already resident in DRAM count toward the target, so no promotion is needed when the existing DRAM share already reaches $r$. The actual promotion volume is further bounded by the eviction-candidate filter above, allowing \textsc{Saria} to balance reload bandwidth against current DRAM contention.

\subsection{Coordinated Multi-Source Reload}
\label{sec:design-coordinator}

The source balance in \S\ref{sec:design-promotion} assumes that DRAM and GDS make progress concurrently.  However, we find that simply enabling overlap does not ensure concurrent progress.
Figure~\ref{fig:coordinator-traffic-evaluation} shows that SSD traffic remains low during the initial DRAM burst and reaches its sustained rate only after DRAM traffic subsides.  A disk-only completion tail therefore persists despite concurrent submission, leaving GPU ingress underutilized.

\textsc{Saria} therefore coordinates the two sources by pacing DRAM traffic while GDS work is active.
The coordinator consumes each request's reload plan from \S\ref{sec:design-prefix}, tracks its remaining DRAM and SSD work, and shares a per-GPU ingress budget across concurrent traffic.  Let $B_{\mathrm{DRAM}}$ denote the standalone DRAM reload bandwidth.  While SSD work is runnable, \textsc{Saria} limits the aggregate DRAM rate to approximately
\begin{equation}
  B_{\mathrm{DRAM}}^{\mathrm{pace}} =
  \min\left(B_{\mathrm{DRAM}},
  B_{\mathrm{GPU}} - B_{\mathrm{disk}}\right),
  \label{eq:dram-pacing}
\end{equation}
reserving sufficient ingress bandwidth for GDS to sustain $B_{\mathrm{disk}}$.
The coordinator enforces this rate by bounding individual DRAM submissions and spacing them according to the available budget. When no SSD work is runnable, DRAM returns to its unthrottled path.

\begin{figure}[t]
  \centering
  \includegraphics[width=0.98\linewidth]{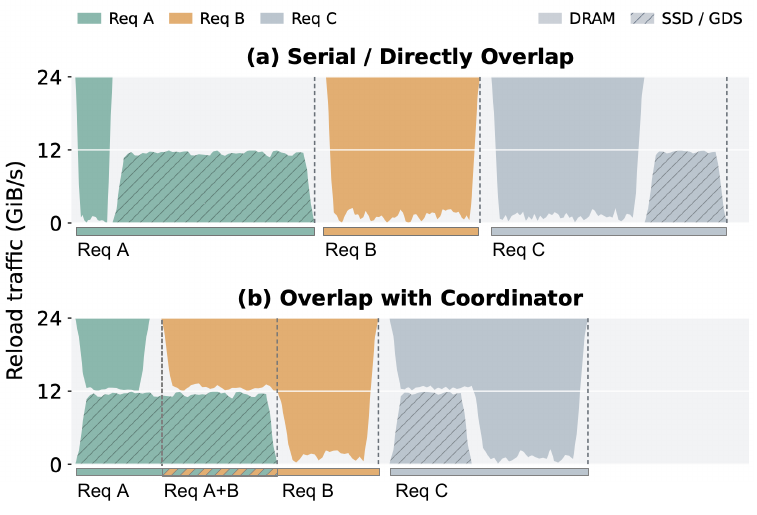}
  \caption{
  Illustration of multi-source reload with and without transfer coordination.
  }
  \label{fig:coordinator-read-traffic}
\end{figure}

Figure~\ref{fig:coordinator-read-traffic} illustrates the coordinated schedule.
Once Request A finishes its DRAM portion, residual ingress advances the DRAM-only Request B while A continues reading from SSD.
When A's SSD transfer finishes, B uses the full DRAM rate.
Request C overlaps its DRAM and SSD transfers; after its SSD portion completes, its remaining DRAM work also runs at full rate.

The coordinator is configured using estimated DRAM, GDS, and GPU-ingress bandwidths. Inaccurate settings can cause bandwidth underutilization or transfer contention, degrading reload performance. When GDS alone saturates GPU ingress bandwidth, coordination offers no benefit and its overhead may reduce performance.





\section{Evaluation}
\label{sec:evaluation}

\subsection{Experimental Setup}
\label{sec:evaluation-setup}

\stitle{Models and testbeds}
Our primary testbed has one RTX PRO 6000 Blackwell Server Edition GPU with 96~GB of GDDR7 memory, two AMD EPYC 9534 CPUs, 256~GB of host memory, and two 1.92-TB PCIe 5.0 NVMe SSDs. We evaluate released BF16 checkpoints of Qwen3-8B with 36 full-attention layers and Qwen3.5-9B with 8 full-attention and 24 gated-delta-network (GDN) layers.


Admission and metadata experiments use a secondary testbed with a 24-GB A30 GPU and one PCIe 5.0 NVMe SSD. To accommodate its HBM while preserving the attention structures, we construct Qwen3-1.7B with 20 full-attention layers and Qwen3.5-5.6B with 5 full-attention and 15 GDN layers. These names denote our scaled-down configurations rather than released checkpoints; both execute real forward passes with randomly initialized BF16 weights, following~\cite{orca,jenga,mooncake}.


\begin{table}[t]
  \centering
  \small
  \setlength{\tabcolsep}{3pt}
  \caption{Selected workloads after filtering and truncation. Minimum prefill tokens denote the total number of prompt tokens that still require computation with an unlimited-capacity cache.}
  \label{tab:workloads}
  \begin{tabular*}{\columnwidth}{@{\extracolsep{\fill}}lrrrr@{}}
    \toprule
    Dataset & Sessions & Requests & \shortstack{Total prompt\\tokens ($10^6$)} & \shortstack{Minimum prefill\\tokens ($10^6$)} \\
    \midrule
    TraceLab & 350 & 6,431 & 538.26 & 36.55 \\
    WEKA & 170 & 5,995 & 930.25 & 26.43 \\
    \bottomrule
  \end{tabular*}
\end{table}

\begin{figure*}[t]
  \centering
  \includegraphics[width=\linewidth]{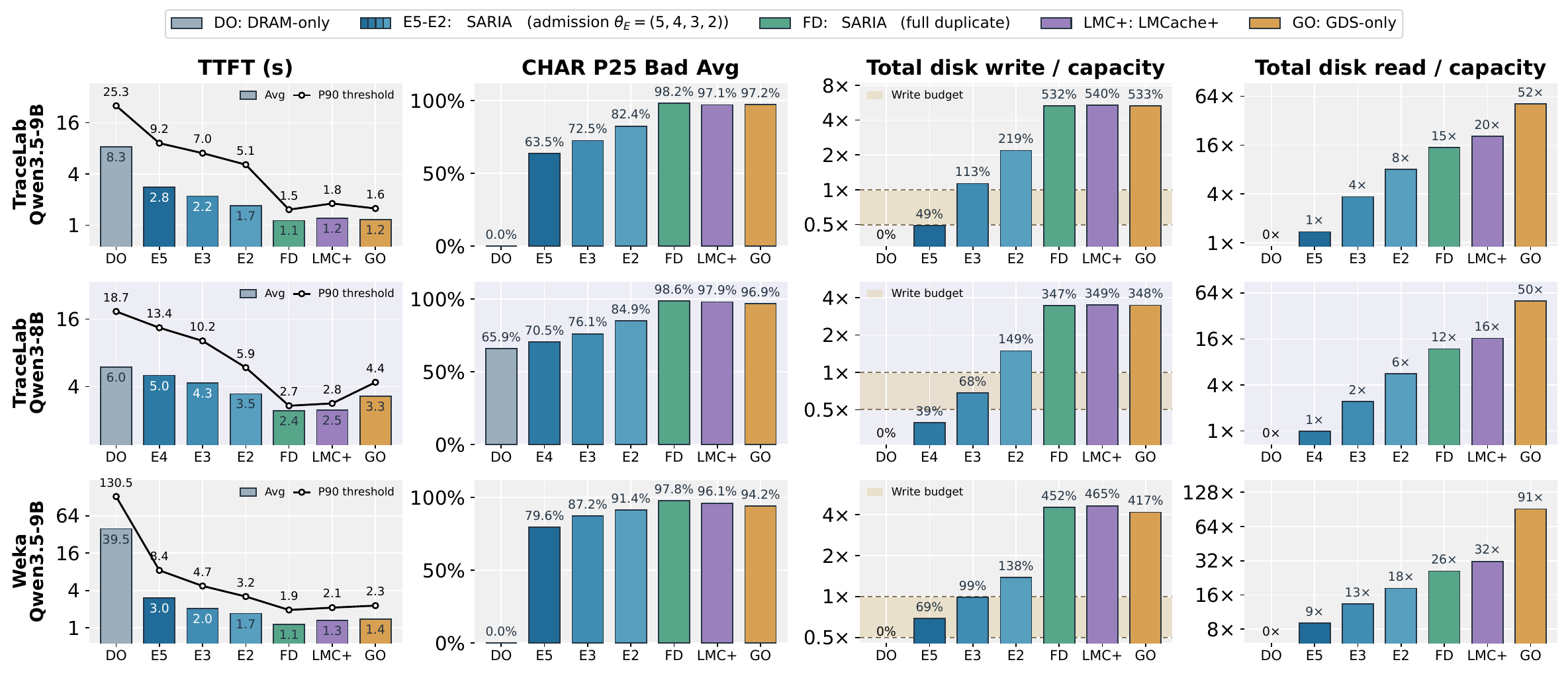}
  \caption{End-to-end latency, cache coverage, and SSD traffic on TraceLab and WEKA. TTFT bars show means and black lines show P90. SSD I/O is normalized by SSD cache capacity. The pale band in the third column spans the 12--24-hour write-budget limits ($0.5$--$1.0\times$ capacity) at 1 DWPD; writes below $0.5\times$ meet both limits.}
  \label{fig:eval-main}
\end{figure*}

\begin{figure*}[t]
  \centering
  \includegraphics[width=\linewidth]{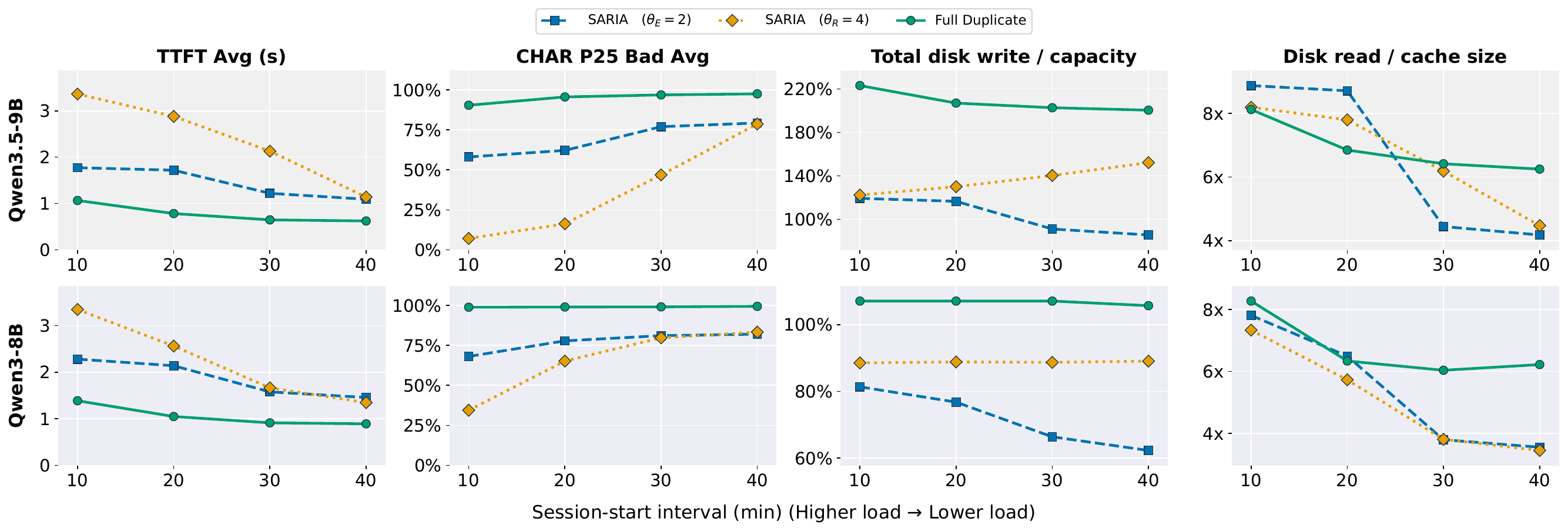}
  \caption{
  Cache behavior under varying serving pressure. A fixed set of 100 sessions is replayed with session-start windows ranging from 10 to 40 minutes, while admission thresholds remain unchanged across all settings.
  }
  \label{fig:eval-pressure}
\end{figure*}

\stitle{Workloads}
We evaluate prefill serving using coding-agent workloads from TraceLab traces of Claude Code and Codex~\cite{tracelabDataset}, and SemiAnalysis WekaTrace (WEKA) traces of Claude Code~\cite{semianalysisWeka}.
Neither dataset includes prompt contents, so we reconstruct deterministic synthetic inputs preserving recorded reuse. TraceLab records request token counts and timestamps; its prefixes are reconstructed from within-session request lengths. WEKA inputs use recorded 64-token block identities scoped by source trace and model.

We select TraceLab sessions by start time and WEKA streams by dataset order. Each WEKA stream forms an independent agent session, including requests after prefix resets. Both workloads use a 450K-token context limit and 24-hour truncation (Table~\ref{tab:workloads}), with mean within-session request intervals of $112.1$ and $111.1$~s, respectively.


We apply the replay method from \S\ref{sec:observations-and-motivation} to both datasets, using a default 12-hour session-start window with Gamma-distributed gaps ($k=1$) following~\cite{servegen}.

Replay permits at most one active request per session. Successors follow recorded idle intervals. We allow four requests in flight by default; the cumulative ablation and promotion sweep use one. Requests stop after the first output token to isolate prefill serving. The logical replay clock advances at 6,500 computed prefill tokens/s; reported latencies use wall-clock time.

\stitle{Baselines and cache configuration}
We implement \textsc{Saria} on LMCache MP and integrate Phoenix as its GDS backend.
Our implementation includes three optimizations to the LMCache MP data path. First, GDS control operations are asynchronously pipelined to reduce control-path overhead. Second, when DRAM allocation fails during HBM offload, the cache evicts existing entries and retries the allocation, allowing offload to continue under memory pressure. Third, following Tutti~\cite{tutti}, we separate SSD read and write execution to reduce I/O interference and improve aggregate throughput.

Two LMCache-based baselines use optimized Phoenix GDS backends~\cite{lmcache}: GDS-only stores KV exclusively on SSD; LMCache+ uses full duplicate DRAM--SSD caching with per-object metadata, serial tier loading, and no promotion. We also run \textsc{Saria} as DRAM-only (SSD disabled) and Full Duplicate (selective admission disabled, metadata and reload mechanisms retained).

Unless otherwise specified, DRAM-containing configurations receive budgets of 1M tokens for DRAM and 10M tokens for SSD. Hybrid budgets use full-attention token equivalents; GDN checkpoints still consume physical capacity and contribute to I/O. Dual-SSD configurations stripe each object across both drives.

We use eager execution with HBM prefix caching disabled. External cache blocks contain 512 tokens for dense models and 1,024 for hybrid models; GDN checkpoints are retained every 16,384 tokens. By default, \textsc{Saria} uses promotion ratio $r=0.4$ and shared-ingress/GDS/DRAM bandwidth estimates of 48/20/48~GB/s (24/10/24~GB/s on A30). The promotion recency threshold is fixed at $P_{\tau}=60$~s on the replay clock. Experiment-specific overrides accompany the results.

\stitle{Metrics}
TTFT measures wall-clock time from submission to vLLM through the first output token; cache-transfer time spans the first external-cache lookup through vLLM-observed load completion, excluding prefill computation. We report mean and 90th-percentile (P90) TTFT in the main comparison. Latency P25 bad averages are means of the largest 25\% of finite latency samples.

Cache hit achievement rate (CHAR) is the ratio of actual cache-hit block occurrences to those achievable with unlimited capacity; 100\% means all available reuse is captured. CHAR P25 bad avg recomputes this ratio over the combined block counts of the quarter of eligible requests with the lowest per-request CHAR. Normalized SSD reads and writes divide cumulative completed payload bytes by configured SSD cache capacity, including GDN state and transfers across both drives.

\subsection{Main Results}
\label{sec:evaluation-main}

\stitle{End-to-End Performance}
Figure~\ref{fig:eval-main} evaluates how \textsc{Saria} balances latency against SSD write budgets and read bandwidth.  Using a standard-endurance TLC SSD rated at 1 DWPD~\cite{kioxiaCD8PR}, we allocate endurance in proportion to the configured cache capacity $C$.  The 12-hour session-start window and 24-hour trace truncation give write budgets of $0.5C$ and $C$, respectively, marked by the shaded band.

On WEKA with Qwen3.5-9B, an eviction threshold $\theta_E=3$ (E3) stays within the one-day write budget, reducing writes by 78.2\% relative to Full Duplicate (FD) at a 78.4\% increase in mean TTFT.  It still reduces mean TTFT by about 94.9\% relative to DRAM-only.  E2 further improves latency but exceeds the one-day write limit.

TraceLab exhibits the same trade-off but requires different thresholds.  For Qwen3.5-9B, E3 reduces writes by 78.7\% relative to FD but exceeds the one-day budget by 13.3\%.  Its mean and P90 TTFT are 73.7\% and 72.2\% lower than DRAM-only.  Raising the threshold to E5 meets even the 12-hour write limit while retaining mean and P90 TTFT reductions of 66.5\% and 63.7\%.  For Qwen3-8B, E3 meets the one-day budget with 80.3\% fewer writes than FD, while reducing mean and P90 TTFT by 27.8\% and 45.3\% relative to DRAM-only.  The threshold needed to meet a write budget thus depends on both workload and model.

DRAM-assisted reload also reduces SSD read traffic: FD issues 27.2\% fewer read bytes than LMCache+ for both models on TraceLab and 18.1\% fewer for Qwen3.5-9B on WEKA.  Together, admission and reload management balance latency, write-budget consumption, and SSD read demand.

\stitle{Performance under Varying Pressure}
We examine whether a fixed eviction threshold can adapt SSD persistence to changing serving load.  We replay the earliest 100 sessions with their starts distributed over 10-, 20-, 30-, or 40-minute windows.  All settings retain the same 1,960 requests without truncation and use 0.5M-token DRAM, 5M-token SSD capacity.  Smaller windows increase concurrent demand.  Figure~\ref{fig:eval-pressure} compares eviction-based admission ($\theta_E=2$), reuse-gated admission ($\theta_R=4$), and full duplicate caching, with admission thresholds fixed across all load levels.

Eviction-based admission reduces persistence as serving pressure falls.  Expanding the session-start window from 10 to 40 minutes reduces SSD writes by 28.3\% for Qwen3.5-9B and 23.5\% for Qwen3-8B, while improving P25 bad CHAR.  Mean TTFT also varies over a narrower range than with reuse-gated admission for both models.  With a fixed threshold, eviction history thus adapts persistence to load: higher pressure makes more blocks qualify for SSD retention, while lighter load allows DRAM to cover more reuse and conserves SSD writes.

\begin{figure}[t]
  \centering
  \includegraphics[width=\linewidth]{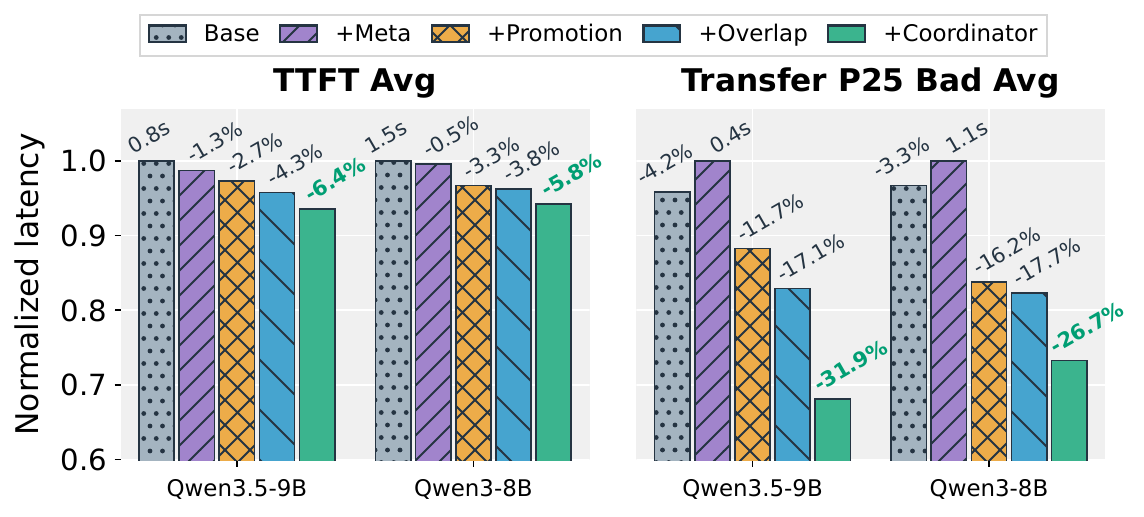}
  \caption{
  Cumulative ablation of \textsc{Saria}'s reload mechanisms under full duplicate caching. Mean TTFT is normalized to Base, while mean transfer time over the slowest 25\% of requests is normalized to +Meta.
  }
  \label{fig:eval-factor}
\end{figure}

\stitle{Ablation Study for SARIA Design}
We isolate the reload mechanisms using the 350-session workload with full duplicate caching and one in-flight request.  Starting from per-object metadata and serial tier loading (Base), we cumulatively add unified metadata and mixed-source lookup (+Meta; \S\ref{sec:design-prefix}), partial DRAM promotion (\S\ref{sec:design-promotion}), tier overlap, and coordination (\S\ref{sec:design-coordinator}).  Figure~\ref{fig:eval-factor} normalizes mean TTFT to Base and P25 bad transfer time to +Meta.  Since Meta increases cache hits and the transferred volume, Base transfer time is shown only for reference.

The complete configuration reduces mean TTFT by 6.4\% for Qwen3.5-9B and 5.8\% for Qwen3-8B relative to Base, and P25 bad transfer time by 31.9\% and 26.7\% relative to +Meta.  Adding the coordinator to the overlap-enabled configuration further reduces this tail metric by 17.8\% and 11.0\%, respectively.  Source placement and transfer scheduling contribute together, with the largest relative gains in cache-transfer latency.

\subsection{Micro Experiments}
\label{sec:evaluation-ablations}

\begin{figure}[t]
  \centering
  \includegraphics[width=1\linewidth]{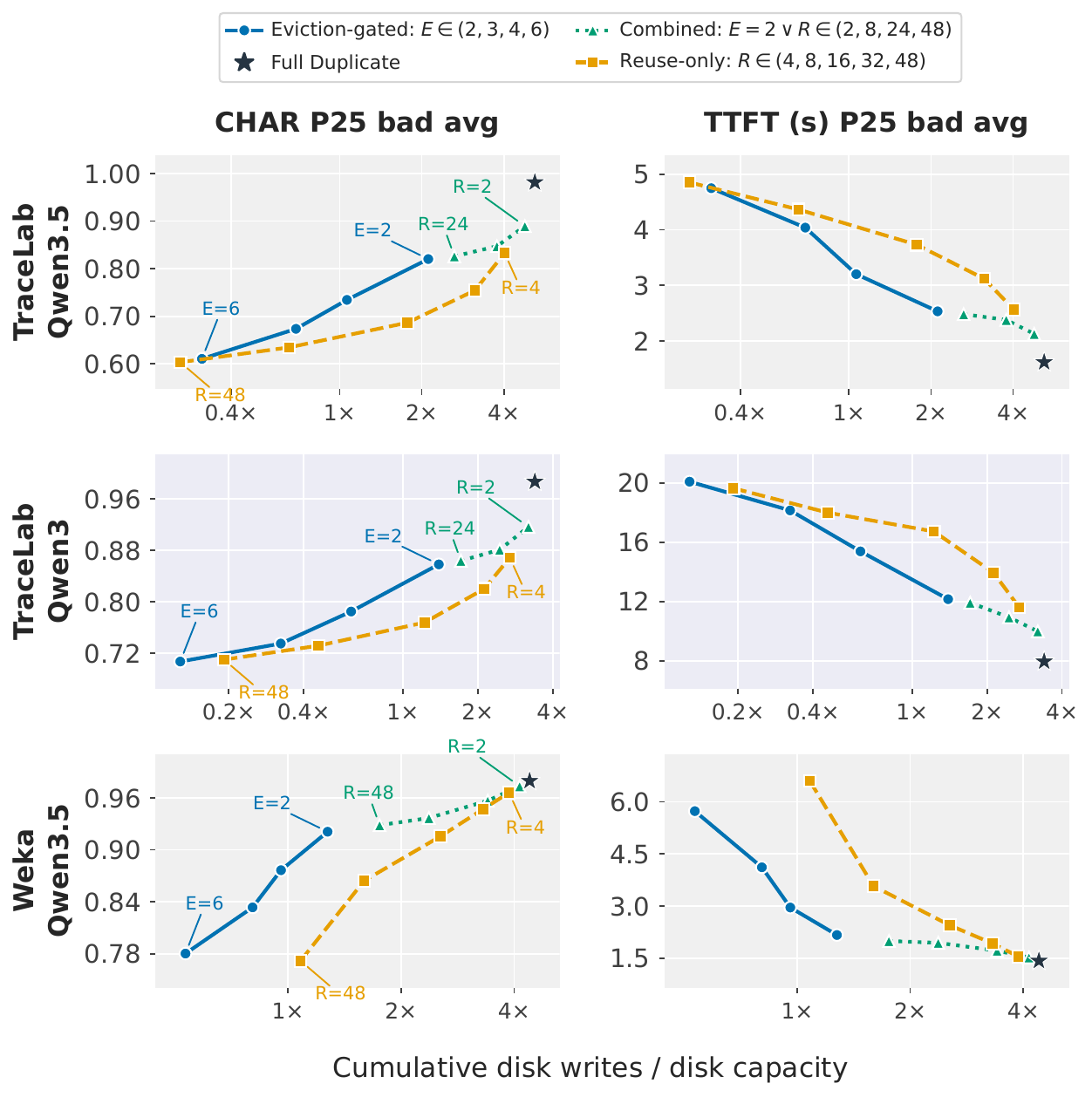}
  \caption{
  Comparison among eviction-based, reuse-based, and combined admission policies on TraceLab and WEKA using the A30 testbed.
  }
  \label{fig:eval-admission}
\end{figure}

\stitle{Disk Admission Trade-offs}
Figure~\ref{fig:eval-admission} evaluates the two admission signals on A30 using TraceLab and WEKA.  We vary the eviction threshold, the reuse-only threshold, and their combination while retaining the same cache capacities and reload mechanisms.  The displayed eviction thresholds are $E\in\{2,3,4,6\}$; reuse-only thresholds are $R\in\{4,8,16,32,48\}$; combined settings fix $E=2$ and use $R\in\{2,8,24\}$, with an additional $R=48$ setting evaluated only on WEKA.  Full duplicate caching provides a reference with no admission filtering.

Eviction-based admission reaches comparable reuse coverage with substantially fewer writes.  On TraceLab, reuse-only $R=4$ requires about 1.9 times as many writes as E2 for similar P25 bad CHAR on both models.  Adding the reuse branch with $R=2$ to E2 further improves coverage for popular contexts at greater write cost.  Together, the signals expose a range of write--coverage trade-offs, with lower P25 bad TTFT as more reusable KV cache is retained.

\stitle{Unified Metadata and Reload Planning}
We evaluate unified context metadata and mixed-source reload planning by enabling or disabling them together under E2 admission on A30.  The comparison holds promotion, overlap, coordination, and cache capacities fixed.

\begin{figure}[t]
  \centering
  \includegraphics[width=\linewidth]{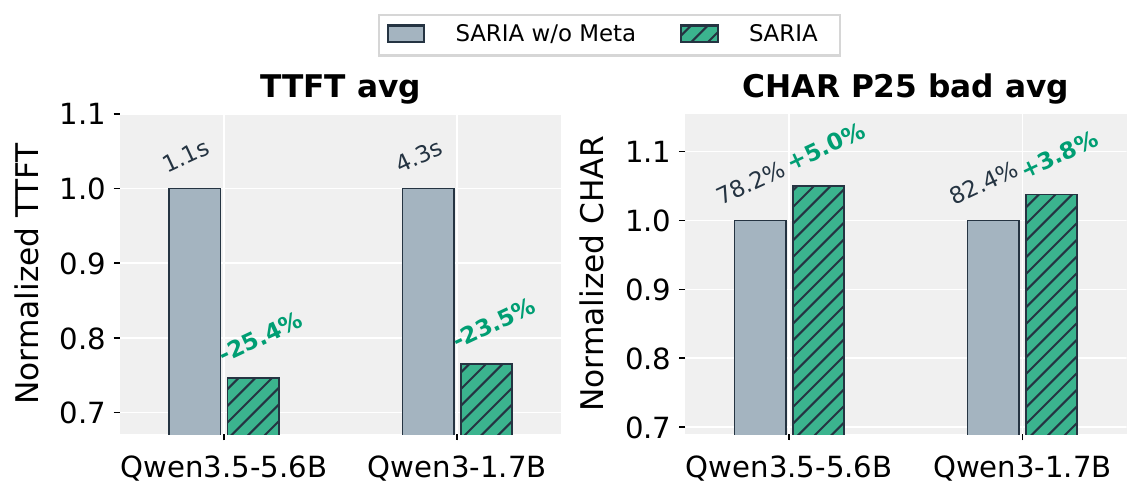}
  \caption{
  Comparison of unified context metadata and mixed-source lookup under E2 admission. All values are normalized to the configuration with both mechanisms disabled.
  }
  \label{fig:eval-metadata}
\end{figure}

Figure~\ref{fig:eval-metadata} shows that enabling these mechanisms reduces mean TTFT by 25.4\% for Qwen3.5-5.6B and 23.5\% for Qwen3-1.7B, while improving P25 bad CHAR for both models.

The loss of reuse without these mechanisms stems in part from LMCache's tier-prefix lookup, which cannot extend a prefix with DRAM-only blocks after switching to SSD.  Selective admission can produce DRAM--SSD--DRAM prefixes, so omitting the final segment forces recomputation of existing KV cache blocks.  Unified reload planning recovers these interleaved hits and reduces redundant prefill (\S\ref{sec:design-prefix}).

\begin{figure}[t]
  \centering
  \includegraphics[width=\linewidth]{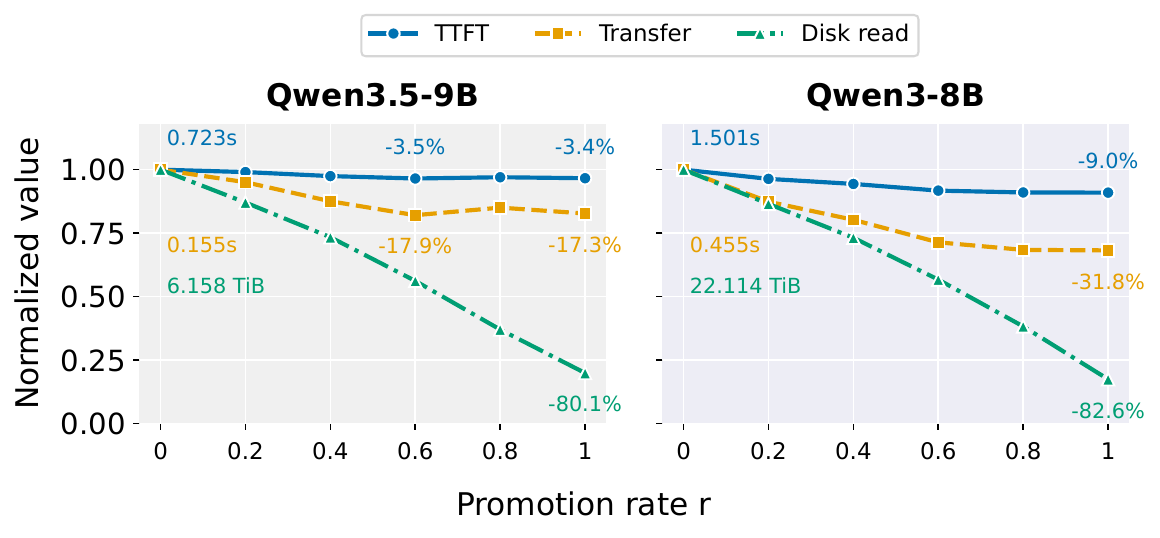}
  \caption{
  Promotion-ratio sweep under full duplicate caching with coordinated reload. Metrics are normalized to the zero-promotion configuration ($r=0$).
  }
  \label{fig:eval-promotion}
\end{figure}

\stitle{Lazy Partial DRAM Promotion}
To isolate the benefit of promotion, we fix full duplicate caching and coordinated reload and vary the target DRAM share $r$, using the 350-session workload with one in-flight request.  Figure~\ref{fig:eval-promotion} normalizes mean TTFT, mean cache-transfer time, and SSD read volume independently to their $r=0$ values.  Both $r=1$ points use later implementation snapshots and are included as references; comparisons below use $r\leq0.8$ from a common snapshot.

Promotion reduces SSD traffic and accelerates subsequent reloads.  For Qwen3.5-9B, $r=0.6$ reduces SSD reads by 43.8\%, mean transfer time by 17.9\%, and mean TTFT by 3.5\%.  For Qwen3-8B, $r=0.8$ yields reductions of 61.7\%, 31.6\%, and 9.0\%, respectively.  On Qwen3.5-9B, increasing $r$ from 0.6 to 0.8 further reduces SSD reads but does not improve latency.  This agrees with the bandwidth balance in \S\ref{sec:design-promotion}: DRAM can complement SSD bandwidth without holding the entire reload set, so additional read savings need not shorten reloads.

\begin{figure}[t]
  \centering
  \includegraphics[width=\linewidth]{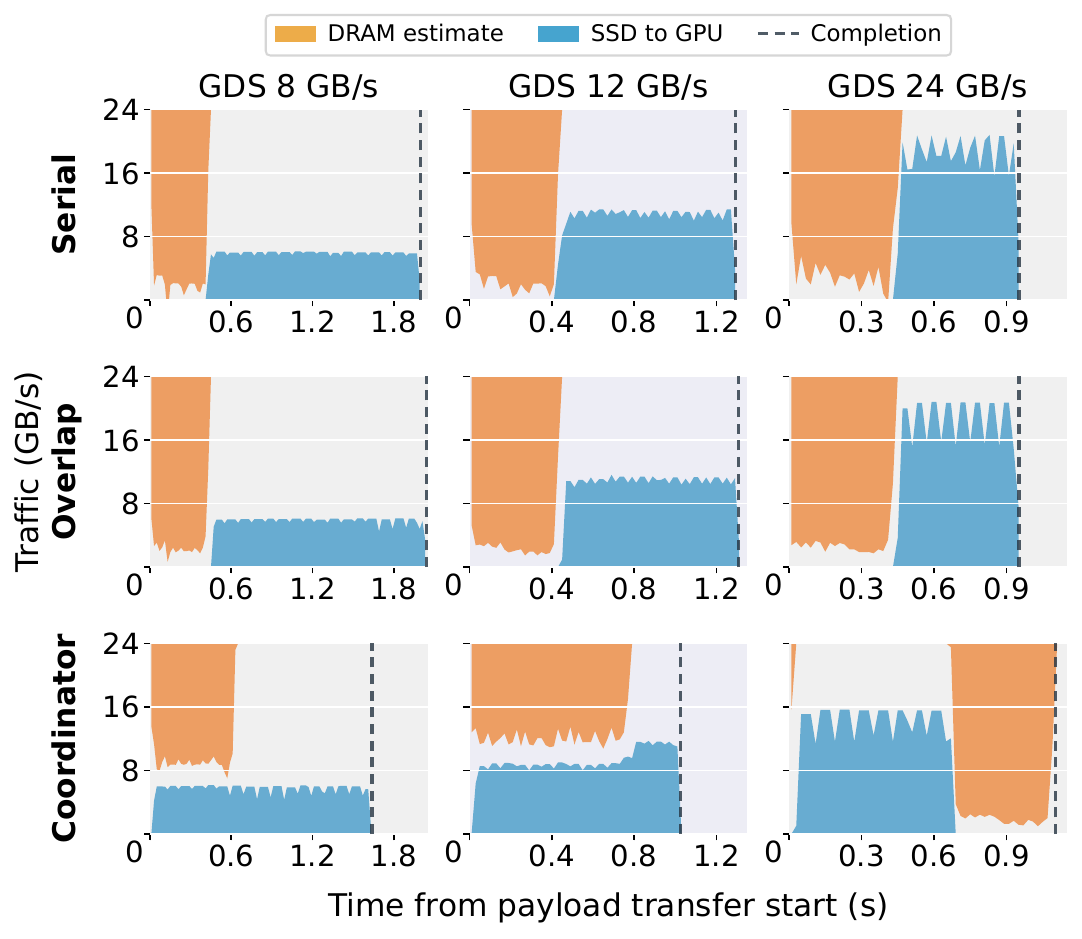}
  \caption{
  Comparison of coordinated and uncoordinated reload traffic on the A30 testbed with equal DRAM and SSD payloads under different hardware.
  }
  \label{fig:coordinator-traffic-evaluation}
\end{figure}

\stitle{Coordinated Multi-Source Reload}
We isolate transfer contention on A30 with a single synthetic request using the Qwen3-8B KV layout and no model computation.  The request reloads 135,168 cached tokens, split equally between DRAM and SSD (9.28125~GiB per tier).  Figure~\ref{fig:coordinator-traffic-evaluation} compares serial loading, uncoordinated overlap, and coordinated overlap reload for a slower SSD, a faster SSD, and two striped SSDs.  The coordinator uses a 24~GB/s shared-ingress budget and respective GDS allowances of 8, 12, and 23.9~GB/s.  SSD traffic is reconstructed from DMA events; DRAM traffic is estimated from PCIe receive traffic after subtracting SSD traffic and normalized to the known DRAM payload.  Each plot contains one measured reload in 20-ms bins, timed from payload submission.

For both single-SSD configurations, uncoordinated overlap leaves SSD progress low during the initial DRAM burst and provides no improvement over serial loading.  Coordination allows both sources to progress, reducing completion time by 18.0\% on the slower SSD and 20.7\% on the faster SSD relative to serial loading.  With two SSDs, however, coordination increases completion time by 15.8\% relative to serial loading.  The disk allowance leaves little bandwidth for DRAM, while coordinator overhead prevents SSD traffic from reaching that allowance.  Coordination therefore helps under limited SSD bandwidth, but its allowances must match effective path bandwidths (\S\ref{sec:design-coordinator}).

\section{Related Work}
\label{sec:related-work}

\stitle{KV cache management}
SGLang~\cite{sglang} organizes reusable prefixes with radix trees, while HiCache~\cite{sglangHiCache} further uses access counts to filter HBM-to-DRAM backups. Unified Radix Cache~\cite{unifiedRadixCache} extends prefix management to hybrid-attention models, and Marconi~\cite{marconi} introduces checkpoint-aware caching for recurrent states. SAECache~\cite{saecache} uses eviction feedback to estimate token reuse value. These techniques improve cache organization and admission within memory hierarchies, but do not explicitly address SSD endurance. \textsc{Saria} instead uses reuse and eviction history across DRAM residencies to determine which KV cache blocks warrant persistence on SSD.

\stitle{SSD-backed KV caching}
Mooncake~\cite{mooncake,mooncakeStoreSSD}, LMCache~\cite{lmcache}, and CachedAttention~\cite{cachedattention} extend KV caching beyond GPU memory using host DRAM and SSDs. HiCache reloads SSD-resident KV cache through DRAM and supports full promotion, whereas FlexKV~\cite{flexkvGDSReload} and Dynamo KVBM~\cite{dynamoKVBMReload} use GDS to reload SSD hits directly into GPU memory without DRAM promotion. Tutti~\cite{tutti} exploits SSD parallelism and an optimized I/O stack to approach DRAM-backed cache performance. Unified KV Pooling~\cite{unifiedkvpooling} partitions KV cache according to device bandwidth, while HiFC~\cite{hifc} combines GDS with pseudo-SLC regions. Strata~\cite{strata} and Swarm~\cite{swarm} improve storage I/O scheduling and multi-SSD placement. In contrast, \textsc{Saria} targets deployments with limited SSD endurance and parallelism, using partial DRAM promotion and coordinated mixed-source reloads to reduce dependence on the bandwidth-limited SSD path without excessive DRAM competition.

\stitle{SSD Admission and Write Endurance}
Flashield~\cite{flashield} reduces flash writes by admitting general-purpose key--value objects based on their observed reads and updates, while CacheSack~\cite{cachesack} and Baleen~\cite{baleen} similarly optimize flash-cache admission. These policies target general storage workloads rather than the session-level reuse structure of agent serving. \textsc{Saria} exploits this structure by distinguishing reuse already covered within a DRAM residency from reuse that extends across DRAM residencies. Repeated eviction and reuse therefore provide an admission signal that reflects both session reuse and cache competition, complemented by a reuse threshold for highly popular prefixes.

\section{Conclusions}
\label{sec:conclusions}



SSD-backed KV caching faces two practical challenges: excessive writes rapidly consume SSD endurance, while limited SSD parallelism constrains reload bandwidth. We present \textsc{Saria}, a hierarchical KV caching system that addresses these challenges through \emph{Persistence-Aware SSD Admission}, which uses reuse and eviction history to avoid unnecessary SSD writes; \emph{Lazy Partial DRAM Promotion}, which retains selected SSD-hit KV cache blocks in DRAM for subsequent reuse while limiting DRAM competition; and coordinated multi-source reload, which schedules DRAM and GDS transfers to efficiently share GPU ingress bandwidth. Across TraceLab-derived workloads and both full-attention and hybrid-attention models, \textsc{Saria} reduces SSD writes by 57.1--58.9\% compared with full duplicate caching while reducing mean TTFT by 42.3--79.7\% compared with DRAM-only caching.





\bibliographystyle{plain}
\bibliography{main}

\end{document}